\documentstyle[preprint,aps,draft]{revtex}

\begin{document}

\title{Anisotropic Exchange in LiCuVO$_4$ probed by ESR}
\author{H.-A.~Krug~von~Nidda$^a$, L.~E.~Svistov$^{a,b}$, M.~V.~Eremin$^{a,c}$,
R.~M.~Eremina$^{a,d}$, A.~Loidl$^a$, V.~Kataev$^{e,d}$,
A.~Validov$^d$, A.~Prokofiev$^{f}$, W.~A{\ss}mus$^{f}$}

\address{$^{a}$ Experimentalphysik V, Elektronische Korrelationen und Magnetismus,
Institut f\"{u}r Physik, Universit\"{a}t Augsburg, 86135 Augsburg,
Germany\\ $^b$ A. V. Shubnikov Institute of Crystallography, RAS,
117333 Moscow, Russia
\\ $^c$ Kazan State University, 420008 Kazan, Russia \\
$^d$ E. K. Zavoisky Physical-Technical Institute, 420029 Kazan, Russia \\
$^{e}$ II. Physikalisches Institut, Universit\"{a}t zu K\"{o}ln, 50937
K\"{o}ln, Germany
\\ $^{f}$ Physikalisches Institut,
Johann-Wolfgang-Goethe-Universit\"{a}t, D-60054 Frankfurt, Germany}

\date{\today}
\maketitle

\begin{abstract}
We investigated the paramagnetic resonance in single crystals of
LiCuVO$_4$ with special attention to the angular variation of the absorption
spectrum. To explain the large resonance linewidth of the order of 1 kOe, we
analyzed the anisotropic exchange interaction in the chains of edge-sharing
CuO$_6$ octahedra, taking into account the ring-exchange geometry of the
nearest-neighbor coupling via two symmetric rectangular Cu-O bonds. The
exchange parameters, which can be estimated from theoretical considerations,
nicely agree with the parameters obtained from the angular dependence of the
linewidth. The anisotropy of this magnetic ring exchange is found to be much
larger than it is usually expected from conventional estimations which neglect
the bonding geometry. Hence, the data yield the evidence that in copper oxides
with edge-sharing structures the role of the orbital degrees of freedom is
strongly enhanced. These findings establish LiCuVO$_4$ as one-dimensional
compound at high temperatures.



\end{abstract}

\smallskip

\section{\protect\smallskip INTRODUCTION}

In the past years experimental and theoretical studies of
low-dimensional quantum magnetism of transition-metal (TM) oxides
have received much attention. In particular, the discovery of a
spin-Peierls transition in the one-dimensional Heisenberg
antiferromagnet CuGeO$_3$ by Hase \textit{et al.} \cite{hase93} in
1993 triggered an intensive search for such a ground state in TM
oxides with spin $S=1/2$ ions like Cu$^{2+}$ or V$^{4+}$. In this
contents a lot of interesting low-dimensional compounds was
investigated. However up to now, no second inorganic spin-Peierls
system was found, whereas this singlet-dimerization of spin pairs
is often observed in organic chain compounds. Even in the
spin-ladder system NaV$_2$O$_5$, which exhibits a comparable
magnetic susceptibility like CuGeO$_3$, the ground state turned
out to be driven by charge order \cite{thalmeier98}. In spite of
this failure, recent research has brought into light another
important aspect of magnetism of one-dimensional $S = 1/2$ chains
made of TM ions. A closer look at the properties of magnetic
interactions in a number of Cu-based oxides revealed a
surprisingly strong deviation from isotropic Heisenberg magnetism,
which at first glance is not expected at all due to the quenching
of the orbital angular momentum of Cu$^{2+}$ by the crystal field
\cite{boehm98,fong99,ammerahl00,kataev01}. Indeed, the magnetic
anisotropy due to the orbital degrees of freedom plays a minor
role in the two-dimensional cuprates, such as the parent compound
of the high-T$_{\rm c}$ superconductors La$_2$CuO$_4$. The
important difference between two- and one-dimensional Cu-oxides
arises from the fact that in the former the exchange interaction
is usually mediated via a single 180$^{\circ}$ Cu-O-Cu bond,
whereas in the latter the coupling often proceeds via two
symmetric almost rectangular Cu-O-Cu bonds. The leading isotropic
superexchange interaction ${\cal H}_{\rm iso}=\sum J_{ij}^{\rm
iso} {\bf S}_i{\bf S}_j$ is reduced very strongly when going from
180$^{\circ}$ to 90$^{\circ}$, as it follows from the
Goodenough-Kanamori-Anderson rules \cite{goodenough55}. However,
it turns out that this reduction does not concern the anisotropic
corrections to ${\cal H}_{\rm iso}$. This is for example justified
by the observation of a very large spin-wave gap in Li$_2$CuO$_4$
\cite{boehm98}, a large size of the ordered moment in
Ca$_2$Y$_2$Cu$_5$O$_{10}$ \cite{fong99}, or extremely anisotropic
magnetic order and a very broad electron-spin-resonance (ESR)
linewidth in the paramagnetic state of
La$_5$Ca$_9$Cu$_{24}$O$_{41}$ \cite{ammerahl00,kataev01}. The
subject of the present paper, LiCuVO$_4$, exhibits Cu chains
comparable to CuGeO$_3$ and to the other Cu oxides mentioned
above, but it has been discussed controversially as one- or
two-dimensional antiferromagnet \cite{gonzalez94,yamaguchi96}.

LiCuVO$_4$ crystallizes within an inverse spinel structure
$AB_2$O$_4$, which is orthorhombically distorted due to the
cooperative Jahn-Teller effect of the Cu$^{2+}$-ions (electronic
configuration $3d^9$). The nonmagnetic V$^{5+}$ ions are placed on
the $A$ positions in oxygen tetrahedra, whereas Li$^+$ and
Cu$^{2+}$ occupy the $B$ positions in oxygen octahedra
\cite{blasse65}. Both LiO$_6$ and CuO$_6$ octahedra form
independent chains along the $a$ and $b$ directions, respectively,
which themselves built up a stack of independent planes of Li and
Cu rods along the $c$ direction \cite{okeefe77}. The detailed
crystallographic data and a picture of the structure of LiCuVO$_4$
are given by Lafontaine \textit{et al.} \cite{lafontaine89}. The
chains consist of edge-sharing octahedra with two nearly
rectangular Cu-O-Cu superexchange bonds between every two Cu ions.
Within the planes, neighboring chains are connected with each
other by VO$_4$ tetrahedra forming Cu-O-V-O-Cu superexchange
bonds. As from Goodenough-Kanamori-Anderson rules the $90^{\circ}$
Cu-O-Cu superexchange is expected to be ferromagnetic, but becomes
antiferromagnetic for slight deviations of the bond angle from
$90^{\circ}$ as well as due to possible side-group effects
\cite{geertsma96}, the relative strength of the different exchange
paths is still unknown.

Due to this uncertainty, the magnetic susceptibility of LiCuVO$_4$
has been interpreted in different ways
\cite{gonzalez94,yamaguchi96,blasse65,vasilev99}. With decreasing
temperature the susceptibility deviates from its high-temperature
Curie-Weiss law $\chi\propto(T-\Theta_{\rm CW})^{-1}$ with
$\Theta_{\rm CW} = -15$~K. It develops a first broad maximum
close to $T_{\rm max} = 28$~K, which is followed by a second
sharper peak around 3~K, and finally drops down below $T_{\rm N}=
2.3$~K. Originally the broad maximum was ascribed to
two-dimensional antiferromagnetism between the Cu chains within
one plane, and the sharp peak to the onset of three-dimensional
order \cite{gonzalez94,blasse65}. Later, one-dimensional
antiferromagnetism has been proposed to explain the behavior
around 28~K followed by two-dimensional correlations near 3~K
\cite{yamaguchi96} and finally three-dimensional order at 2.3~K
\cite{vasilev99}.

Recently, ESR has been carried out in polycrystalline LiCuVO$_4$
by Vasil'ev \textit{et al.} \cite{vasilev01}. The spectra reveal
the typical pattern due to a uniaxial anisotropy of the resonance
field, which allows to determine the $g$ values $g_{\parallel}$
and $g_{\perp}$ and the linewidths $\Delta H_{\parallel}$ and
$\Delta H_{\perp}$ for the magnetic field applied parallel or
perpendicular to the local symmetry axis, respectively. The $g$
values, which are found to amount $g_{\parallel}=2.25$ and
$g_{\perp}=2.08$ at high temperatures, are typical for Cu$^{2+}$
in tetragonally distorted octahedral environment. As the long axis
of the CuO$_6$ octahedra is oriented along the crystallographic
$c$ direction, $g_{\parallel}$ can be identified with $g_c$. Below
100~K the anisotropy $g_{\parallel}-g_{\perp}$ slowly increases up
to a factor 2 at $T_{\rm N}$ compared to the value at room
temperature. The linewidth is of the order of 1~kOe revealing an
anisotropy as well with $\Delta H_{\parallel}/\Delta H_{\perp}
\approx 1.5$. Its temperature dependence exhibits a minimum near
16~K and increases monotonously with negative curvature on
increasing temperature. To lower temperatures the linewidth
diverges on approaching the onset of magnetic order.

In a previous work \cite{kegler01} we independently performed ESR
experiments on oriented powder samples, where the single
crystalline grains were aligned by a magnetic field along their
$c$ axis. The ESR spectra consist of a single Lorentzian line
showing an orientation dependence in accordance to the results
obtained by Vasil'ev \textit{et al.} in randomly distributed
powder. The integrated intensity of the ESR line was found in good
agreement with the static susceptibility. Concerning the resonance
linewidth, we discussed the strength and the possible type of
anisotropy of the spin-spin coupling in LiCuVO$_4$. We estimated
the contributions of dipole-dipole (DD), anisotropic exchange
(AE), and Dzyaloshinsky-Moriya (DM) interaction to $\Delta H$
following the work of Yamada \textit{et al.} in CuGeO$_3$
\cite{yamada96}, which exhibits a comparable linewidth. From this
estimation, only the DM interaction, which is of first order in
the spin-orbit coupling and thus brings the largest anisotropic
correction to ${\cal H}_{\rm iso}$, yielded the appropriate
contribution to the line broadening, whereas the contributions
from AE and DD interaction were found to be one and two orders of
magnitude too small, respectively. However, considering the
symmetry of the unit cell, the DM interaction should be zero,
because every two DM vectors along the Cu chains cancel each
other. In addition, a recent theoretical study of ESR in
one-dimensional magnets with the DM interaction claims that its
contribution to the linewidth should be in any case of the same
level as that of the symmetric anisotropic term
\cite{choukroun01}.

In the present paper, we take the advantage of the measurements in
single crystals of LiCuVO$_4$, which allows us to study very
accurately the angular dependence of the ESR spectrum. In
particular, the angular variation of the linewidth provides us
with a new insight into the magnetic interactions in this
compound. We reanalyze the anisotropic exchange within the Cu
chains using the ideas, which were recently proposed to explain
the width of the ESR signal in the Cu-O chains of the so called
telephone-number compound La$_{14-x}$Ca$_{x}$Cu$_{24}$O$_{41}$
\cite{kataev01}. The situation in that compound is quite similar
to LiCuVO$_4$, as every two Cu ions in the chain are connected
with each other by two oxygen ions, giving rise to a strongly
anisotropic ring exchange, which is one order of magnitude larger
than we estimated before. We show that in the high-temperature
limit the linewidth can be mainly described in terms of this
strong symmetric anisotropic exchange and can rule out the
presence of the DM interaction. Our results emphasize that an
adequate description of the magnetic properties of low-dimensional
Cu oxides requires a detailed study of the anisotropic couplings
which depend sensitively on a particular bonding geometry.

\section{Anisotropic exchange in the Cu chains}

\subsection{ESR linewidth and g factor}

We consider a system of exchange-coupled spins ${\bf S}_i$ with an
effective spin Hamiltonian given by

\begin{equation}\label{Hamilton}
{\cal H}_{\rm eff}= \sum\limits_{i>j}\left[J_{ij}^{\rm iso}\left({\bf
S}_{i}{\bf S}_{j}\right) + {\mathbf S}_i {\mathbf J}_{ij}
{\mathbf S}_j \right] + \sum\limits_{i} \mu_{\rm B} {\mathbf H}
{\mathbf g} {\mathbf S}_i
\end{equation}

where the scalar $J_{ij}^{\rm iso}$ denotes the isotropic exchange
between two spins $i$ and $j$, and the tensor ${\mathbf J}_{ij}$
is the anisotropic interaction due to symmetric exchange and
dipole-dipole interaction. The last term describes the Zeeman
splitting of the spin states in an external magnetic field
${\mathbf H}$ with gyromagnetic tensor ${\mathbf g}$ and Bohr
magneton $\mu_{\rm B}$. In the case of LiCuVO$_4$ it is enough to
take into account the interaction between neighboring Cu spins
within the Cu-O chains ($i=j+1\rightarrow J_{ij}^{\rm iso} \equiv
J$), where all Cu places are equivalent \cite{remark-nnn}. In the
crystallographic coordinate system, where the Cartesian
coordinates $(x,y,z)$ are chosen parallel to the $(a,b,c)$ axes
of the orthorhombic unit cell, both the $g$ tensor with
components $(g_a,\ g_b,\ g_c)$ and the tensor of the anisotropic
interaction $(J_{xx},\ J_{yy},\ J_{zz})$ are diagonal.

Following Anderson and Weiss \cite{anderson53}, in the case of
strong exchange narrowing the ESR linewidth $\Delta H$ (in Oe) is
obtained from the second moment $M_2$ of ${\cal H}_{\rm eff}$ via
the relation

\begin{equation}\label{dh}
\Delta H = \frac{\hbar}{g\mu_{\rm B}\omega _{\rm ex}}M_{2}
\end{equation}

where $\omega_{\rm ex} \approx J/\hbar = \mid J_{xy,xy} \mid
/\hbar$ is the so called exchange frequency due to the
super-exchange coupling of the ground-states $|xy>$ (see next
subsection below) and

\begin{equation}\label{g}
g = \sqrt{g_{c}^{2}\cos^{2}\theta + (g_{a}^{2}\cos ^{2}\varphi +
g_{b}^{2}\sin ^{2}\varphi)\sin ^{2}\theta}
\end{equation}

is the usual expression for the $g$ factor. Polar angle $\theta$
(with respect to $c$) and azimuth $\varphi$ (with respect to $a$)
give the orientation of the external field in the crystallographic
system. In coordinates $\tilde{x}, \tilde{y}, \tilde{z}$, where
the $\tilde{z}$-axis is defined by the direction of the applied
magnetic field $\textbf{H}$, the second moment $M_{2}$ due to
anisotropic exchange is given by \cite{soos77,pilawa97}

\begin{equation}\label{M2AE}
M_{2}=2\frac{S(S+1)}{3}\left\{
f_{1}(2\tilde{J}_{zz}-\tilde{J}_{xx}-\tilde{J}_{yy})^{2}
+f_{2} \cdot 10(\tilde{J}_{xz}^{2}+\tilde{J}_{yz}^{2})
+f_{3}[(\tilde{J}_{xx}-\tilde{J} _{yy})^{2}+4\tilde{J}
_{xy}^{2}]\right\}
\end{equation}

The factor 2 appeared due to the summation over next neighbors.
The symbols $f_{1}$, $f_{2}$, and $f_{3}$ denote the so called
spectral-density functions, as introduced by Pilawa
\cite{pilawa97}. $f_{1}$ corresponds to the secular part, whereas
$f_{2}$ and $f_{3}$ correspond to nonsecular parts.
$\tilde{J}_{\alpha \beta}$ are anisotropy parameters in notations
of reference \cite{soos77}. On transformation to the
crystallographic coordinates ($x,y,z$), we obtain the angular
dependence (see appendix).

Usually the resonance field of the paramagnetic resonance is obtained
from the first moment of the spin Hamiltonian. In our case we
find for ${\bf H}||c$:

\begin{equation}\label{reso}
\hbar \omega _{c} = g_{c}\mu_{\rm B}H_{\rm
res}\{1-\frac{\chi_{c}(T)}{g_{c}\mu_{B}} (2J_{zz}-J_{xx}-J_{yy})\}
\end{equation}

where $\chi _{c}(T)$ is the susceptibility per one copper site
along the $c$ axis. One gets the analogous expressions for ${\bf
H}||a$ and ${\bf H}||b$ by changing $(x,y,z;c)$ into $(y,z,x;a)$
and $(z,x,y;b)$, respectively. For the case that one-dimensional
short-range-order effects in the chain are important we find,
following to the method described by Nagata and Tazuke
\cite{nagata72}

\begin{equation}\label{reso2}
\hbar \omega_{c}=g_{c}\mu _{B}H\{1-\frac{N(T)}{\mid J\mid}
(2J_{zz}-J_{xx}-J_{yy})\}
\end{equation}

and analogous expressions for $a$ and $b$ direction. Here $N(T)$
is given by

\begin{eqnarray}
N(T)&=&\frac{1}{10x}\left[
\frac{2+ux}{1-u^{2}}-\frac{2}{3x}\right] \nonumber \\
u&=&x+\coth (-\frac{1}{x}) \nonumber \\
x&=&\frac{k_{B}T}{2\mid J\mid S(S+1)} \nonumber
\end{eqnarray}

We note that at the specified choice of anisotropy parameters
(axial symmetry along the chain) our expressions convert to those
which are listed by Nagata and Tazuke.

\subsection{Estimation of the anisotropic exchange parameters}

As it was described by many authors (see for example
\cite{bleaney52,yosida96}), the anisotropic exchange (AE) appears
due to a virtual hopping process of an electron or hole between
the ground-state orbital $d_{xy}$ and the excited states
$d_{x^2-y^2}$, $d_{yz}$, and $d_{zx}$ in combination with
spin-orbit coupling. Here the notation "$xy$" and "$x^2-y^2$" is
interchanged with respect to the conventional usage, because we
have chosen the coordinate system along the crystal axes, where
the $x$ and $y$ axes are rotated by 45$^{\circ}$ with respect to
the Cu-O bonds in the chains. For Cu$^{2+}$ with $3d^9$ electron
configuration there is a single hole with $S = 1/2$ in the ground
state. In our case the most effective path for hopping is from
the $d_{xy}$ orbital of one Cu ion ($j$) via an oxygen $p$ orbital
to the excited $d_{x^2-y^2}$ state of the neighboring Cu ion
($i$), as it is explained in Fig. \ref{AE}. The excited
$d_{x^2-y^2}$ state is connected to the ground state $d_{xy}$ of
the same Cu ion ($i$) via spin-orbit coupling $\xi({\mathbf S}_{i}
\cdot {\mathbf L}_{i})$. Here ${\mathbf S}_{i}$ and ${\mathbf
L}_{i}$ denote the spin and orbital momentum of the Cu ion ($i$),
respectively. There is only one matrix element $<x^{2}-y^{2}\mid
l_{z}\mid xy>=-2i$ of the $l_{z}$ operator, which connects the
$d_{x^{2}-y^{2}}$ and $d_{xy}$ states. Therefore this process
contributes to $J_{zz}$ only, yielding the AE parameter

\begin{equation}
J_{zz}^{\rm AE} = 2\left( \frac{\xi}{E_{x^2-y^2}-E_{xy}}\right)
^{2}J_{x^{2}-y^{2},xy}=\frac{1}{32}\left(g_{zz}-2\right)
^{2}J_{x^{2}-y^{2},xy}
\end{equation}

with the energies $E_{xy}$ and $E_{x^2-y^2}$ of the ground state
and excited state, respectively. The AE parameters $J_{xx}^{\rm
AE}$ and $J_{yy}^{\rm AE}$ are given by

\begin{equation}
J_{xx}^{\rm AE} = \frac{1}{2}\left(
\frac{\xi}{E_{zx}-E_{xy}}\right)^{2}J_{zx,xy}
=\frac{1}{8}\left(g_{xx}-2\right) ^{2}J_{zx,xy}
\end{equation}

\begin{equation}
J_{yy}^{\rm AE} = \frac{1}{2}\left(
\frac{\xi}{E_{yz}-E_{xy}}\right)^{2}J_{yz,xy}
=\frac{1}{8}\left(g_{yy}-2\right)^{2}J_{yz,xy}
\end{equation}

Here $J_{\alpha,xy}$ (with $\alpha = x^2-y^2$, $yz$, $zx$) denotes
the exchange-interaction parameter between one Cu ion in the
ground state and the other Cu ion in the excited $d_{\alpha}$
state. In addition, we have to take into account the
dipole-dipole (DD) interaction

\begin{eqnarray}
J_{zz}^{\rm DD}=\mu_{\rm B}^{2}\frac{g_{zz}^{2}}{R^{3}},\
J_{xx}^{\rm DD}=\mu_{\rm B}^{2}\frac{g_{xx}^{2}}{R^{3}},\
J_{yy}^{\rm DD}=-2\mu_{\rm B}^{2}\frac{g_{yy}^{2}}{R^{3}}
\end{eqnarray}

where $R$ denotes the distance between neighboring Cu ions within
the chains.

In the case under consideration, there are two bridging oxygen
ions (see Fig. \ref{AE}). As it was shown in Ref.
\cite{eremin82,voronkova83}, in such a geometry quantum
interference between different exchange paths in the Cu-O
plaquette strongly intensifies the ferromagnetic super-exchange
coupling parameter $J_{x^{2}-y^{2},xy}$ with respect to the
parameter $J_{xy,xy}$ for the ground state of the copper ions.
This was experimentally confirmed by ESR measurements of
metal-organic complexes \cite{voronkova83}, where an unusually
large anisotropic exchange parameter $J_{zz}^{\rm AE}$ was found
for Cu-Cu pairs connected via two bridging oxygen ligands. In the
context of spin chains in copper oxides, such an effect was
discussed in Ref. \cite{yushankhai99,tornow99} as ring exchange.

In reference \cite{voronkova83} the parameter $J_{x^{2}-y^{2},xy}$
was estimated as -330~cm$^{-1}$. If we assume that this value is
relevant for the case of LiCuVO$_4$ too, then with
$g_{zz}-2=0.33$ (see Ref. \cite{kegler01} and the next section
below), we expect $J_{zz}^{\rm AE}\approx -2$~K. The parameter
$J_{zx,xy}$ has a ferromagnetic character (i.e $< 0$) and it is
small with respect to $J_{x^{2}-y^{2},xy} $, therefore
$J_{xx}^{\rm AE}$ is not important. Using the
Goodenough-Kanamori-Anderson rules, one can conclude that the
parameter $|J_{yz,xy}|>|J_{zx,xy}|$. Unfortunately, it is not easy
to estimate, but definitely $J_{yy}^{\rm AE}$ is not too small
with respect to $J_{yy}^{\rm DD}$. Substituting the intra-chain
Cu-Cu distance $R=2.786$~${\rm \AA}$ into the dipole-dipole
terms, one should finally expect $J_{xx}\approx J_{xx}^{\rm
DD}\approx 0.12$~K, $J_{yy} \approx J_{yy}^{\rm AE}+J_{yy}^{\rm
DD} \gg -0.24$~K and $J_{zz}=J_{zz}^{\rm AE}+J_{zz}^{\rm DD}
\approx -1.8$~K.

Using these results we can estimate the resonance linewidth in
LiCuVO$_4$ from Eqns. \ref{dh} and \ref{M2AE}. With the
ground-state super-exchange integral $J = J_{xy,xy} = -2
\Theta_{\rm CW} = 30$~K and the AE parameter $J_{zz}$ estimated
above, one obtains $\Delta H \approx 1.5$~kOe. As we will see in
the next section, this value is in very good agreement with the
linewidth experimentally observed in single crystals of
LiCuVO$_4$.

\section{Experimental Results and Discussion}

The starting materials for synthesis of LiCuVO$_4$ were
Li$_2$CO$_3$ (99.9\%), V$_2$O$_5$ (99.5\%) and CuO (99.5\%).
Single crystals of LiCuVO$_4$ were grown by slow cooling of a 40-mol.\%
solution of LiCuVO$_4$ in a LiVO$_3$ melt. The cooling from 675$^{\circ}$C to
580$^{\circ}$C was carried out with the rate of 0.8$^{\circ}$C/h.
The details were described earlier \cite{prokofiev00}. X-ray
diffraction analysis shows the absence of inclusions of other phases.
The cell parameters derived from Rietveld refinement are $a =
5.645$~${\rm \AA}$, $b = 5.800$~${\rm \AA}$ and $c = 8.747$~${\rm \AA}$.

The ESR measurements were performed with a Bruker ELEXSYS E500
CW-spectrometer at X-band frequency ($\nu \approx 9.35$~GHz),
equipped with continuous gas-flow cryostats for He (Oxford
Instruments) and N$_2$ (Bruker), which allow to cover a
temperature range between 4.2 K and 680 K. For temperatures down
to 1.7 K, we used a cold-finger $^4$He-bath cryostat. The ESR
spectra record the power $P_{\rm abs}$ absorbed by the sample from the
transverse magnetic microwave field as a function of the static
magnetic field $H$. The signal-to-noise ratio of the spectra is
improved by detecting the derivative $dP_{\rm abs}/dH$ with lock-in
technique. The LiCuVO$_4$ single crystals were glued on a
suprasil-quartz rod, which allowed the rotation of the sample
around defined crystallographic axes. The magnetic susceptibility
was measured in a commercial SQUID magnetometer (QUANTUM DESIGN)
within a temperature range $1.8 < T < 400$~K.

In the paramagnetic regime, the ESR spectrum consists of a
single exchange-narrowed resonance line \cite{anderson53} at all
orientations of the magnetic field with respect to the crystallographic
axes.  Figure \ref{spectra} illustrates typical ESR spectra for the magnetic
field applied parallel to the $c$ axis at different temperatures. The parameter
$A$ denotes the relative amplification factor. The resonance is well fitted by
a Lorentzian lineshape with a small contribution $| \alpha | \ll 1$ of
dispersion

\begin{equation}
\frac{dP_{\rm abs}}{dH} \propto \frac{d}{dH}\left[\frac{\Delta H + \alpha
(H-H_{\rm res})}{(H-H_{\rm res})^2 + \Delta H^2} + \frac{\Delta H
+ \alpha (H+H_{\rm res})}{(H+H_{\rm res})^2 + \Delta H^2}\right]
\label{dyson}
\end{equation}

As the linewidth $\Delta H$ is of the same order of magnitude as
the resonance field $H_{\rm res}$ in the present compounds,
equation \ref{dyson} takes into account both circular components
of the exciting linearly polarized microwave field and therefore
includes the resonance at reversed magnetic field $-H_{\rm res}$.

Admixture of dispersion to the absorption signal is usually
observed in metals, where the skin effect drives electric and
magnetic microwave components out of phase in the sample
\cite{barnes81}. Here, we are dealing with an insulator, where the
asymmetry arises from the influence of non-diagonal elements of
the dynamic susceptibility: This effect is often observed in
systems with interactions of low symmetry and sufficiently broad
resonance lines ($H_{\rm res} \approx \Delta H$) \cite{benner83}.

The intensity $I_{\rm ESR}$ of the ESR line was determined from
$I_{\rm ESR} = A_{\rm sig} \cdot \Delta H^2 (1+\alpha)^{0.5}$,
where $A_{\rm sig}$ denotes the amplitude of the ESR Signal
$dP_{\rm abs}/dH$. The ESR intensity measures the spin
susceptibility and follows nicely the static susceptibility
$\chi$ obtained by SQUID measurements, as it is shown in Fig.
\ref{chi}. The susceptibility is in well agreement with previous
experiments \cite{gonzalez94,yamaguchi96,blasse65,vasilev99},
which have been quoted in the introduction. Now, even the slight
discrepancy in the Curie-Weiss temperatures $\Theta_{\rm CW}$,
which was found in powder samples between ESR and SQUID
measurements \cite{kegler01}, can be ruled out as an artefact.
The high-temperature Curie-Weiss law of both data extrapolates to
$\Theta_{\rm CW} = -15$~K.

Figure \ref{temperature} depicts the temperature dependence of
resonance field and linewidth for the magnetic field applied
parallel to all three crystallographic axes. We confined our
measurements to temperatures below 400~K, because we found an
irreversible increase of the resonance linewidth on heating to
higher temperatures. Probably the heating in N$_2$-atmosphere
reduces the oxygen content in the sample. The data are in good
agreement with those obtained in polycrystalline samples \cite{kegler01}.
In addition the single-crystal measurements reveal also an
anisotropy in the $ab$ plane.  The ratio of the linewidth at high
temperature amounts $\Delta H_c : \Delta H_a : \Delta H_b = 2 :
(5/4) : 1$. Below 200~K the temperature dependence of the
linewidth is strongly nonlinear, whereas above 200~K it increases
linear with temperature. The linewidth due to pure spin-spin
relaxation always reaches an asymptotic value $\Delta H(\infty)$
at high temperatures. Assuming the linear behavior above 200~K
due to additional relaxation processes via the lattice, we can
treat the data at 200~K approximately as the asymptotic
high-temperature value.

Figure \ref{dhth} shows the full angular dependence of the
resonance linewidth for the three crystallographic planes $ab$,
$ac$, and $bc$ at 200~K. The solid lines were obtained by the fit
with equation \ref{dh} using the second moment of anisotropic
exchange from equation \ref{M2AE} \cite{fit-remark1}. The fitting
parameters $J_{xx}=0.16$~K, $J_{yy}=-0.02$~K, and $J_{zz}=-1.75$~K
nicely agree with the preliminary theoretical estimation in
section II \cite{fit-remark2}. It turns out that the AE parameter
$J_{zz}$ is indeed the dominant contribution, whereas $J_{xx}$
can be understood in terms of DD interaction, alone. Concerning
the parameter $J_{yy}$, which was difficult to estimate, the
experiment indicates that both DD and AE nearly cancel each
other. Remarkably, the largest anisotropic contribution to the
superexchange amounts to about 6\% of the leading isotropic
coupling $J \approx 30$~K. Hence, it is three times larger than a
conventional estimate of the anisotropy $\Delta J \sim (\Delta
g/g)^2J$, where $\Delta g$ denotes the shift of the $g$ factor
from the spin-only value \cite{moriya60}. In terms of the
linewidth the latter estimate yields a value, which is an order
of magnitude smaller compared to the experimentally observed one
\cite{kegler01}. That comparison of the theoretical predictions
with the experimental data underlines the importance of the
particular bonding geometry for the magnetic anisotropy of
low-dimensional TM oxides \cite{chi-remark}.

With the above estimates of the anisotropic exchange parameters
the appropriate components of the $g$ tensor $g_a = 2.070$, $g_b
= 2.095$ and $g_c = 2.313$ were obtained from the simultaneous
fit of equation \ref{reso} to the temperature dependence of the
$g$ values, which is shown in figure \ref{gT}. Using the
Curie-Weiss susceptibility with $\Theta_{\rm CW} = -15$~K, the fit
curves reasonably reproduce the experimental data down to 25~K.
The weak linear decrease of the $g$ values with increasing
temperature, which seems to be superposed, may arise from the
slight change of the contribution $\alpha$ of dispersion with
increasing linewidth (about 10\% within the whole temperature
range), which itself affects the resonance field (see Eq.
\ref{dyson}). At lower temperatures ($T < 25$~K), the
experimental susceptibility is reduced with respect to the
Curie-Weiss law, and therefore the data do not follow the
divergence of the theoretical curve. We also tried to fit the
$g$~values with equation \ref{reso2}, which should better account
for one-dimensional fluctuations, but this demands an even
stronger divergence, which is not observed experimentally at all.
This indicates the importance of the inter-chain couplings for
the analysis of the low-temperature behavior.

For a complete analysis of the temperature dependence of
the linewidth further theoretical effort is necessary. Not only the
second moment but also the exchange frequency usually changes with
temperature. Moreover, the inter-chain couplings have to be taken
into account, as well.

\section{CONCLUSION}
To summarize, we presented angular dependent ESR measurements in
single crystals of LiCuVO$_4$. The anisotropic exchange
interaction within the Cu chains was shown to account for the
large linewidth and was successfully applied to describe its
angular dependence. The anisotropic exchange parameters obtained
from the experimental data well agree with theoretical
estimations both in sign and magnitude. Hence from an ESR point
of view, LiCuVO$_4$ turns out to be one-dimensional at high
temperatures. Notably the large magnetic anisotropy in the spin
chains of this compound can be entirely described in terms of the
anisotropic symmetric exchange only, without invoking the
Dzyaloshinsky-Moriya interaction. Its relatively high value is
due to a significantly enhanced role of the orbital degrees of
freedom in the Cu-O chains with 90$^{\circ}$ bonding geometry. In
this respect the present results may be important for the
controversial discussion about the existence of the
Dzyaloshinsky-Moriya interaction in CuGeO$_3$
\cite{yamada96,choukroun01,pilawa97,oshikawa}, as well as for the
better understanding of the magnetic properties of other
low-dimensional transition-metal oxides \cite{CGO-remark}.

\section*{ACKNOWLEDGEMENTS}
We are grateful to D. Vieweg, M. M\"{u}ller and A. Pimenova for
susceptibility measurements. We thank B. I. Kochelaev
(Kazan-State University) for useful discussions. This work was
supported by the Bundesministerium f\"ur Bildung und Forschung
(BMBF) under contract No. 13N6917 (EKM) and partly by the Deutsche
Forschungsgemeinschaft (DFG) via the Sonderforschungsbereich (SFB)
484 and DFG-project No. 436-RUS 113/628/0. The work of M. V.
Eremin was partially supported by University of Russia grant
N991327. The work of V. Kataev was supported by the DFG via SFB
608, and that of A. Validov by the Russian Foundation for Basic
Research RFBR (grant N01-02-17533).

\section*{Appendix}

On transformation of the anisotropic exchange parameters
$\tilde{J}_{ij}$ to the crystallographic coordinates ($x,y,z$), we
obtain in equation \ref{M2AE}:

\begin{eqnarray}
[2\tilde{J}_{zz}-\tilde{J}_{xx}-\tilde{J}_{yy}]^{2} &=&
[J_{zz}(3\cos ^{2}\beta -1) + J_{xx}(3\sin ^{2}\beta \cos
^{2}\alpha -1) \nonumber \\ &+& J_{yy}(3\sin ^{2}\beta \sin
^{2}\alpha -1)]^{2} \nonumber
\end{eqnarray}

\begin{eqnarray}
\tilde{J}_{xz}^{2}+\tilde{J}_{yz}^{2} &=& [(J_{xx}\cos ^{2}\alpha
+J_{yy}\sin ^{2}\alpha -J_{zz})\cos \beta \sin \beta + 2J_{yz}\cos
2\beta \sin \alpha ]^{2} \nonumber
\\ &+& [(J_{yy}-J_{xx})\sin \beta \cos \alpha \sin \alpha ]^{2}
\nonumber
\end{eqnarray}

\begin{eqnarray}
[(\tilde{J}_{xx}-\tilde{J}_{yy})^{2}+4\tilde{J}_{xy}^{2}] &=&
[J_{xx}(\cos ^{2}\beta \cos ^{2}\alpha -\sin ^{2}\alpha) +
J_{yy}(\cos ^{2}\beta \sin ^{2}\alpha -\cos ^{2}\alpha) \nonumber
\\ &+& J_{zz}\sin ^{2}\beta ]^{2} + (J_{yy}-J_{xx})^{2}\cos
^{2}\beta \sin ^{2}2\alpha \nonumber
\end{eqnarray}

Here we have taken into account that the $g$ factor is
anisotropic in our case and therefore

\begin{eqnarray}
\cos\alpha =\frac{A}{\sqrt{A^{2}+B^{2}}}, \ \cos\beta
=\frac{C}{\sqrt{A^{2}+B^{2}+C^{2}}} \nonumber
\end{eqnarray}

where

\begin{eqnarray}
A=g_{xx}\sin \theta \cos \varphi, \ B=g_{yy}\sin \theta \sin
\varphi, \ C=g_{zz}\cos \theta \nonumber
\end{eqnarray}

\begin{figure}
\caption{Relevant orbitals to the strong anisotropy of the
superexchange interaction between the Cu spins in the chain via a
two-oxygen bridge. The coordinate $y$ is chosen along the chain
($b$ axis), $z$ is perpendicular to Cu-O tape ($c$-axis).}
\label{AE}
\end{figure}

\begin{figure}
\caption{ESR spectra of LiCuVO$_4$ for ${\bf H}||c$ at different
temperatures. $A$ denotes the relative amplification}
\label{spectra}
\end{figure}

\begin{figure}
\caption{Temperature dependence of the inverse ESR intensity
$1/I_{\rm ESR}$ (left ordinate) and inverse static susceptibility
$1/\chi$ (right ordinate) obtained by SQUID measurements. In both
cases the magnetic field $H$ has been applied parallel to the
crystallographic $a$ axis. The dotted line indicates a Curie-Weiss
law with a Curie-Weiss temperature $\Theta_{\rm CW}=-15$~K. Inset:
ESR-Intensity (left ordinate) and static susceptibility (right
ordinate) below 80 K with $T_{\rm max} \approx 28$~K and $T_{\rm
N} \approx 2.3$~K.} \label{chi}
\end{figure}

\begin{figure}
\caption{Temperature dependence of resonance field (upper frame)
and linewidth (lower frame) for the magnetic field applied
parallel to the three crystallographic axes.}
\label{temperature}
\end{figure}

\begin{figure}
\caption{Angular dependence of the resonance linewidth for three
crystallographic planes. The $x$, $y$ and $z$ direction have been
chosen parallel to $a$, $b$ and $c$ axis, respectively. The solid
lines have been obtained from the fit as described in the text.}
\label{dhth}
\end{figure}

\begin{figure}
\caption{Temperature dependence of the g-value for the magnetic
field applied parallel to the three crystallographic axes. The
solid lines are fit curves as indicated in the text. The $x$, $y$
and $z$ direction have been chosen parallel to $a$, $b$ and $c$ axis,
respectively.}
\label{gT}
\end{figure}


\begin{references}
\bibitem{remark-nnn} We neglect the next-nearest neighbor (NNN)
exchange interaction for the sake of simplicity, since it does not
affect the conclusions of our qualitative discussion. In
principal, it may be noticeable in LiCuVO$_4$ like in CuGeO$_3$.
At least, the same microscopic calculations in Ref.
\cite{yosida96} suggest that, in general, the NNN exchange in the
edge sharing Cu-O chain may be appreciable.
\bibitem{fit-remark1} It is important to note that the
best fit to the data was achieved choosing the spectral functions
$f_{2}/f_{1}=f_{3}/f_{1}=0.91$, where $f_{1} \equiv 1$. In this
case the angular dependence contains not only terms $\propto
\cos^2$, which would be the case only, if $f_{1}=f_{2}=f_{3}$ is
fulfilled, but also a modulation $\propto \cos^4$. Hence, the
nonsecular part is slightly suppressed with respect to the
secular. Obviously this effect should be interesting for a
thermodynamic description.
\bibitem{fit-remark2} As one can see from the
theoretical expressions, we really have only two fitting
parameters, because the isotropic part $J_{xx}+J_{yy}+J_{zz}$
does not contribute to linewidth and shift.
\bibitem{chi-remark} In principle one could also expect an anisotropic
Curie-Weiss temperature due to the estimated anisotropic exchange.
However, $6\%$ anisotropic contribution to the exchange is still
too small to give a measurable effect. For $\Theta_{\rm CW} =
-15$~K this is approximately 1~K, which is smaller than the
experimental error.
\bibitem{CGO-remark} In spite of a comparable anisotropy of g value and
high-temperature linewidth, in CuGeO$_3$ the situation seems to be
more difficult, as the isotropic antiferromagnetic intra-chain
exchange is about 4 times larger than in LiCuVO$_4$ indicating a
substantial deviation from 90$^{\circ}$ Cu-O-Cu bonding geometry.
As it was shown by Tornow at al. \cite{tornow99}, in this case the
anisotropic correction to the superexchange is significantly
reduced as compared to the rectangular bond. Due to this fact one
may expect a weaker magnetic anisotropy in CuGeO$_3$. In
addition, the next-nearest neighbor contribution is more
important.
\end{references}
\end{document}